\documentclass{ws-p9-75x6-50}

\begin{document}

\title{A brief review of Regge calculus in classical numerical
  relativity} 

\author{Adrian P. Gentle and Warner A. Miller}

\address{Theoretical Division (T-6, MS B288), Los Alamos National
  Laboratory, \\Los Alamos NM 87545, USA. E-mail: apg@lanl.gov}


\maketitle

\abstracts{
We briefly review past applications of Regge calculus in classical numerical
relativity, and then outline a programme for the future development of
the field.  We briefly describe the success of lattice gravity in
constructing initial data for the head-on collision of equal mass
black holes, and discuss recent results on the efficacy of Regge
calculus in the continuum limit.
}

It has long been hoped that Regge calculus\cite{regge61}
could provide an exciting
and independent tool in numerical relativity.\cite{wheeler64}
However, it is only in the last decade that this simplicial,
lattice-based approach to gravity has begun to realise its 
potential.\cite{committee,gentle98,gentle99}  In this
brief report we describe recent numerical applications of Regge
calculus, and outline a programme which we believe will aid the
development of the field.

The early development of the Regge approach to lattice gravity focused
heavily on highly symmetric toy models for which there are
corresponding exact solutions of Einstein's equations.  These early
studies typically retained  only a handful of  degrees of freedom in
the lattice, and applications included various cosmological
spacetimes, together with initial data for single and multiple black
holes.  We refer the reader to the bibliography and review by Williams
and Tuckey\cite{williams92} for further details on these important
pioneering efforts. 

\begin{figure}[t]
  \begin{center}
    \begin{tabular}{cc}
      \epsfig{file=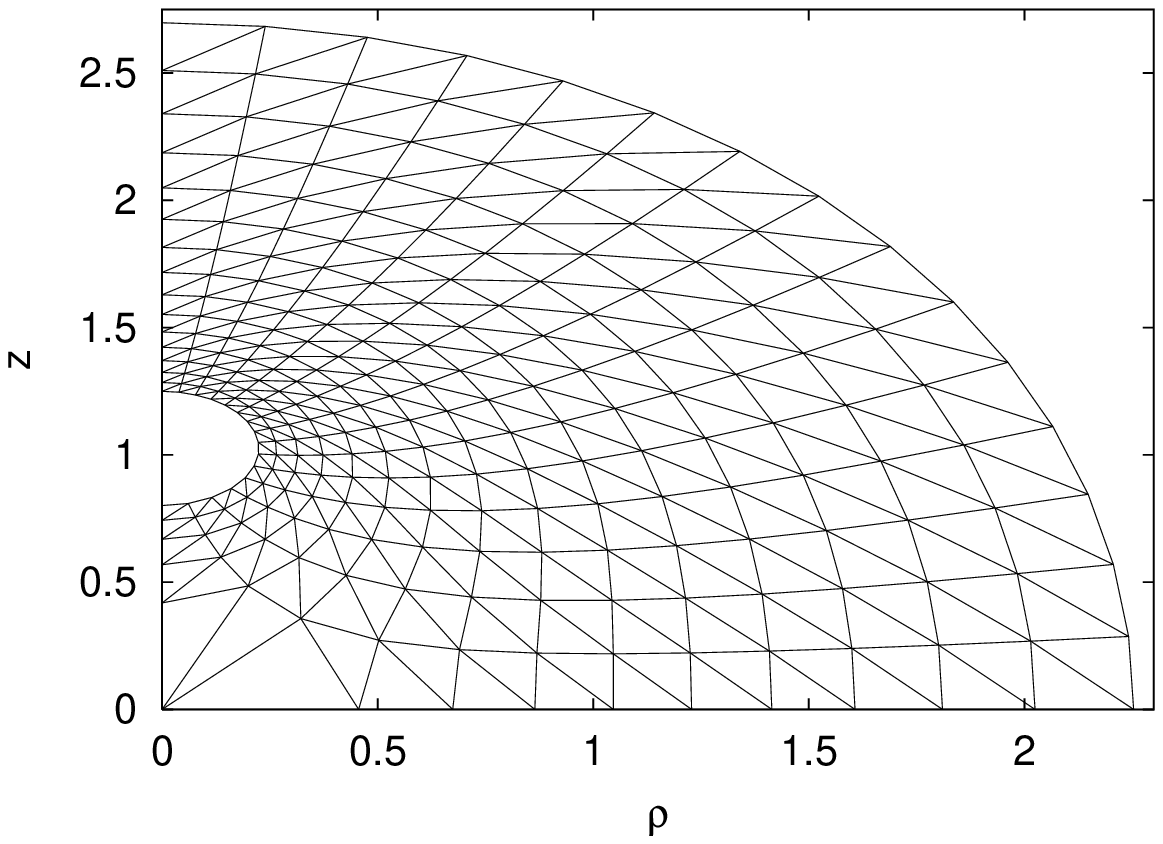, width=0.45\textwidth} &
      \epsfig{file=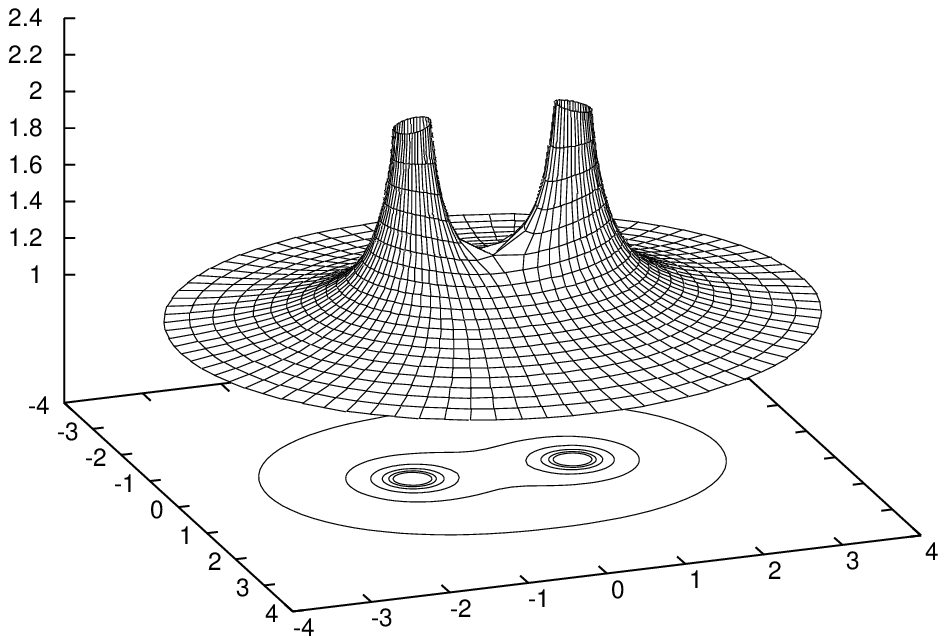, width=0.55\textwidth} \\
      \mbox{\sl(a)} & \mbox{\sl(b)} 
    \end{tabular}
  \end{center}
  \caption{Results from an application of Regge calculus to the
      construction of Misner initial data for the head on collision 
      of non-rotating, equal mass black holes. We show {\em (a)} a
      portion of the lattice, which is based on \v{C}ade\v{z}
      coordinates, and 
      {\em(b)} the conformal factor close to the holes.}
  \label{fig:figa}
\end{figure}

We believe the future development of Regge calculus requires a
rigorous programme of testing and incremental improvements in both our
understanding of the lattice approach and the basic formulation
itself.  To this end, we propose a series of test-bed applications in
which to develop and further our understanding.

The first step of this programme involves a serious examination of
Regge calculus in a generic spherically symmetric setting.  This 
allows the comparison of lattice gravity with more standard finite
difference techniques for the evolution of isolated black hole
spacetimes, and also provides an ideal framework in which to develop a
formalism for the inclusion of matter in Regge calculus.  Work is
currently underway on these issues using a generic simplicial
spherically symmetric lattice.

To complement the spherically symmetric work we are also developing a
generic axisymmetric code, which will be used to study gravitational
radiation on the lattice.  Initial data for Brill waves on
Minkowski and Schwarzschild ``backgrounds'' has already been successfully
constructed.\cite{gentle99a,gentle99b}  In addition, we have also
constructed initial data for the head on collision of two equal mass
black holes\cite{gentle99}; see figure \ref{fig:figa}.  Work is
currently underway on the time evolution of these sets of initial
data. 

Lattice gravity has been successfully benchmarked in three-plus-one
dimensions on the Kasner cosmology\cite{gentle98}; with the insights
gained from the programme  outlined above, the ultimate aim is the
development of the Regge approach to the point where the relative
strengths and weaknesses can be examined in light of more
traditional techniques.  In the short 
term, this involves parallelization of the existing
$(3+1)$-dimensional code, together with the formulation of suitable
lapse and shift conditions.    It is also vital in the longer term
that techniques are 
developed to include generic matter terms in the Regge equations.

Finally, we note that the controversy\cite{brewin95,miller95}
over the convergence 
of Regge calculus to general relativity has been resolved.  The claim
that Regge calculus failed to converge in the continuum limit was
based on the 
observation that the residual of the Regge equations, when evaluated on
carefully interpolated solutions of the Einstein equations, failed to
converge.  We believe that recent work\cite{brewin00} has provided an
explanation of this behaviour which is nevertheless consistent with
second order convergence of the solutions of Regge calculus to the
continuum. 

We believe the future of numerical Regge geometrodynamics is bright.

\end{document}